\documentclass[prb,aps,twocolumn,superscriptaddress,floatfix,showpacs]{revtex4}
\usepackage{graphicx}
\usepackage{color}

\begin{document}

\title{Nature of finite-temperature transition in anisotropic
pyrochlore ${\rm\bf Er_2Ti_2O_7}$}

\author{M. E. Zhitomirsky}
\affiliation{Service de Physique Statistique, Magn\'etisme et Supraconductivit\'e,
UMR-E9001 CEA-INAC/UJF, 17 rue des Martyrs, 38054 Grenoble Cedex 9, France}
\author{P. C. W. Holdsworth}
\affiliation{Laboratoire de Physique, \'Ecole Normale Sup\'erieure de Lyon,
CNRS 69364 Lyon Cedex 07, France}
\author{R. Moessner}
\affiliation{Max-Planck-Institut f\"ur Physik komplexer Systeme, 01187 Dresden, Germany}
\date{February 19, 2014}

\begin{abstract}
We study the finite-temperature transition in a model $XY$ antiferromagnet on
a pyrochlore lattice, which describes the pyrochlore material $\rm Er_2Ti_2O_7$.
The ordered magnetic structure selected by thermal fluctuations is six-fold degenerate.
Nevertheless, our classical Monte Carlo simulations show that the critical
behavior corresponds to the three-dimensional $XY$ universality class.
We determine an additional critical exponent $\nu_6=0.75>\nu$ characteristic
of a dangerously irrelevant scaling variable. Persistent thermal fluctuations
in the ordered phase are revealed in Monte Carlo simulations by the peculiar
coexistence of Bragg peaks and diffuse magnetic scattering, the feature also
observed in neutron diffraction experiments.
\end{abstract}
\maketitle

\begin{figure}[t]
\centerline{
\includegraphics[width=0.9\columnwidth]{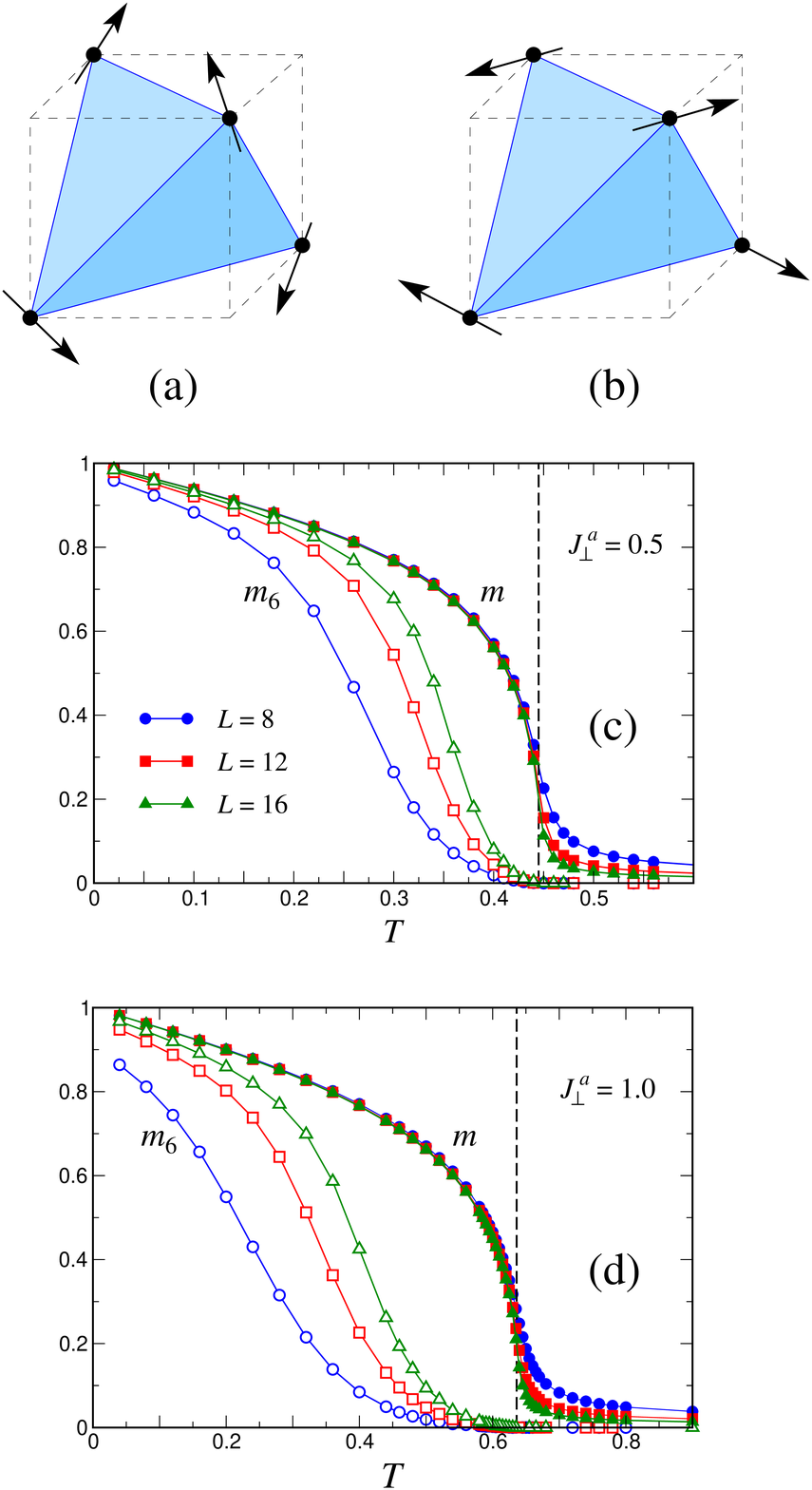}
}
\caption{(Color online)
The noncoplanar magnetic structure of $\rm Er_2Ti_2O_7$, $\psi_2$ state (a)
and the coplanar $\psi_3$ state (b).
Temperature dependence of two order parameters $m$ (full symbols) and
$m_6$ (open symbols) for  $J_\perp^a =0.5$  (c) and $J_\perp^a =1$ (d).
}
\label{fig:OP}
\end{figure}

{\it Introduction.}---%
Geometrically frustrated magnets are widely acknowledged for their exotic disordered
states, which range from classical spin ice \cite{Harris97} to quantum spin liquids
with fractionalized excitations.\cite{Coldea01,Vries09,Han12} Frustrated magnetic
materials also feature various unusual types of order. An incomplete list of such states
includes fractional magnetization plateaus, \cite{Kageyama99,Ueda05,Fortune09}
a partially ordered state, \cite{Stewart04} a valence bond solid, \cite{Tamura06}
and a quantum spin-nematic. \cite{Mourigal12} A recent example of unconventional
magnetic order is provided by $\rm Er_2Ti_2O_7$. This pyrochlore material
undergoes a second-order transition at $T_N\simeq 1.2$~K into a non-coplanar
$k=0$ antiferromagnetic structure, \cite{Champion03,Poole07} see Fig.~\ref{fig:OP}(a).
The peculiarity of this magnetic state stems from the fact that it appears to
be stabilized by quantum and thermal fluctuations, \cite{Zhitomirsky12,Savary12}
a phenomenon known as order by disorder. Despite intense  theoretical and experimental
studies of this and other closely related pyrochlore materials
\cite{Champion03,Poole07,Zhitomirsky12,Savary12,Bramwell94,Champion04,Ruff08,Sosin10,Cao10,
Onoda10,Petrenko11,Ross11,Dalmas12,Bonville13,Hayre13,Wong13,Oitmaa13,Stasiak11,Ross14}
a number of theoretical questions concerning $\rm Er_2Ti_2O_7$ remain tantalizingly open.

One pressing issue to be addressed in the present work is the nature of the transition
at $T_N$. This  is of interest both for the compound in question and as a general
example of fluctuation driven ordering.
The first Monte Carlo simulations on a local axis $XY$ pyrochlore antiferromagnet found a
fluctuation-induced first-order transition. \cite{Bramwell94,Champion04}
In our previous publication we have demonstrated that sufficiently strong anisotropic exchange
interactions modify the transition to a continuous one.
However, the precise critical behavior has not previously been studied. Experimental
measurements\cite{Champion03,Dalmas12} of the order parameter exponent $\beta$
and the specific
heat exponent $\alpha$ indicate that the critical behavior of $\rm Er_2Ti_2O_7$ may be close to the
three-dimensional (3D) $XY$ universality class. In contrast, the ordered antiferromagnetic structure,
Fig.~\ref{fig:OP}(a), breaks only a discrete $Z_6$ symmetry. The emergence of
$U(1)$ symmetry close to the transition point is not entirely surprising. Renormalization-group
theory predicts that a $Z_q$ anisotropy is dangerously irrelevant in 3D for $q\geq 5$. \cite{Blankschtein84,Oshikawa00}
From a numerical point of view, the situation remains less clear. While  early Monte Carlo results
for a clock model \cite{Hove03} and an anisotropic $XY$ model \cite{Lou07}
generally confirm  the scaling conclusions,   two more recent studies of
spin  \cite{Chen10}  and orbital  \cite{Wenzel11}  models with $Z_6$ symmetry find values
of the critical exponents $\beta$ and $\eta$ that are different from those of the 3D $XY$ universality class.
Thus, the critical behavior of models for $\rm Er_2Ti_2O_7$ needs to be clarified from a theoretical standpoint.

Another intriguing observation from neutron diffraction experiments \cite{Ruff08}
is the coexistence, below $T_N$ and in  zero magnetic field, of well-developed magnetic Bragg
reflections and a broad diffuse scattering component.
In the following we use large-scale Monte Carlo simulations to show  that both the 3D $XY$  universality of
the transition and the coexistence of ordered and disordered components in the magnetic neutron scattering
can be naturally understood and explained in the framework of
our minimal spin model for $\rm Er_2Ti_2O_7$. They also add a new twist to the story of
a dangerously irrelevant anisotropy in the 3D $XY$ model.

{\it Model.}---%
The low-temperature magnetic properties of $\rm Er_2Ti_2O_7$ and a number of other pyrochlore
materials are adequately represented by an effective model of interacting Kramers doublets
selected by a strong crystalline electrical field. \cite{Onoda10,Ross11}
The control parameter for this approximation is a ratio of the exchange interaction
$J\sim 0.1$~meV to the crystal-field gap $\Delta \sim 6$~meV.\cite{Champion03}
The resulting pseudo-spin-1/2 Hamiltonian contains only the bilinear terms that are allowed by
the crystal lattice symmetry. \cite{Curnoe08}
Several equivalent representations have been used for this effective Hamiltonian.
Below we use a convenient vector representation:
\begin{eqnarray}
\hat{\cal H} & = & \sum_{\langle ij\rangle}\Bigl\{ J_{zz} S_i^zS_j^z +
J_\perp {\bf S}^\perp_i\!\cdot{\bf S}^\perp_j + J_\perp^a
({\bf S}^\perp_i\!\cdot \hat{\bf r}_{ij})({\bf S}^\perp_j\!\cdot \hat{\bf r}_{ij})
\nonumber \\
& & \phantom{ \sum_{\langle ij\rangle}\Bigl\{} + J_{z\perp}
\bigl[S_j^z({\bf S}^\perp_i\cdot \hat{\bf r}_{ij}) +
S_i^z({\bf S}^\perp_j\cdot \hat{\bf r}_{ji})\bigr] \Bigr\}\,.
\label{H0}
\end{eqnarray}
Here $S_i^z$ is a spin projection onto the local trigonal axis,
${\bf S}^\perp_i$ is the corresponding transverse component, and $\hat{\bf r}_{ij}$
is a unit vector along the bond joining sites $i$ and $j$. Savary {\it et al}.\cite{Savary12}
used inelastic neutron scattering measurements in magnetic field to determine
the exchange parameters corresponding to  $\rm Er_2Ti_2O_7$. Changing between
the two sets of notations for coupling constants, \cite{SM} we obtain
$J_\perp = 0.21$, $J^a_\perp = 0.35$, $J_{zz} = -0.025$, $J_{z\perp} = 0.03$
(all in meV). As $J_{zz}$ and $J_{z\perp}$ are small compared with the in-plane
coupling constants, they can be safely neglected. As a result, one obtains
the minimal model for  $\rm Er_2Ti_2O_7$, in which the anisotropic spin interaction
are restricted to spin components lying in the local $XY$ planes.

{\it Monte Carlo simulations.}---%
We perform Monte Carlo (MC) simulations of the classical model (\ref{H0}) with
$|{\bf S}_i|=1$, $J_{zz} = J_{z\perp} = 0$, and $J_\perp = 1$ and allowing local
$S_i^z$ fluctuations, i.e., considering an effective $XXZ$ model. The
critical behavior is expected to be the same for classical and quantum models.
Note that  the high-temperature series expansion, the only other technique
suitable for numerical investigation of anisotropic pyrochlores, gives
a reasonably accurate estimate for $T_c$ but has been so far unable to predict
the critical properties.\cite{Oitmaa13}

Our MC simulations were done for periodic clusters with $N = 4L^3$ spins for
various values of the control parameter $J_\perp^a$. We use Metropolis single
spin-flip updates restricting the spin motion to increase the acceptance
rate. In addition, micro-canonical over-relaxation steps were added
to accelerate the random walk through phase space. \cite{Creutz87,Zhitomirsky08}
Typically a measurement was
taken after 5 Metropolis steps, followed by 5 over-relaxation sweeps. This hybrid
algorithm performs significantly better than a plain Metropolis algorithm
allowing us to simulate clusters up to $L=30$. Statistical averages and
the error bars were estimated by making $100$ independent cooling runs.

The ordered magnetic state of $\rm Er_2Ti_2O_7$, [$\psi_2$, Fig.~\ref{fig:OP}(a)],
transforms according to the two-component $E$ ($\Gamma_5$) irreducible
representation of the tetrahedral point group $T_d$.
A competing coplanar state $\psi_3$ is shown in Fig.~\ref{fig:OP}(b).
The two states form a basis of the $E$-representation  such that
$\psi_2 \sim (2z^2-x^2-y^2)$ and $\psi_3 \sim (x^2-y^2)$.
Accordingly, any lowest-energy spin configuration may be linearly decomposed into
\begin{equation}
m_x = \frac{1}{N}\sum_i {\bf S}_i \cdot \hat{\bf x}_{n} \,, \quad
m_y = \frac{1}{N}\sum_i {\bf S}_i \cdot \hat{\bf y}_{n} \,,
\end{equation}
where the orthogonal axes $\hat{\bf x}_{n}\perp \hat{\bf y}_{n}$ on each site of a tetrahedron
coincide with spin directions for the two states in Fig.~\ref{fig:OP}.
Our MC simulations measure statistical averages of $m =  (m_x^2 + m_y^2)^{1/2}$ and
\begin{equation}
m_6 = [(m_x+im_y)^6 + (m_x-im_y)^6]/(2m^5)\,.
\end{equation}
The total order parameter $m$ gives the sublattice magnetization, whereas the clock-type order
parameter $m_6= m\cos 6\theta$ distinguishes between $\psi_2$ and $\psi_3$  states: it is positive
for the six non-coplanar $\psi_2$ states, $\theta_k = \pi k/3$, $k=0,\ldots,5$,  and
negative for the coplanar $\psi_3$ structures with  $\theta_k = \pi (1+2k)/6$.

Figures \ref{fig:OP}(c) and \ref{fig:OP}(d) show the temperature dependence of the two order parameters
for $J^a_\perp = 0.5 $ and 1. In each case the transition temperatures,
denoted by vertical dashed lines were obtained from the crossing points of
the Binder cumulants $U_L = \langle m^4\rangle/\langle m^2\rangle^2$ for different
cluster sizes $L$. We find the clock order parameter $m_6$ to be positive  confirming
selection of the $\psi_2$ state by thermal fluctuations.
\cite{Champion04,Zhitomirsky12} However, $m_6$ is strongly suppressed
near $T_c$ and grows as the system size increases at fixed $T<T_c$.

\begin{figure}[t]
\centerline{
\includegraphics[height=0.4\columnwidth]{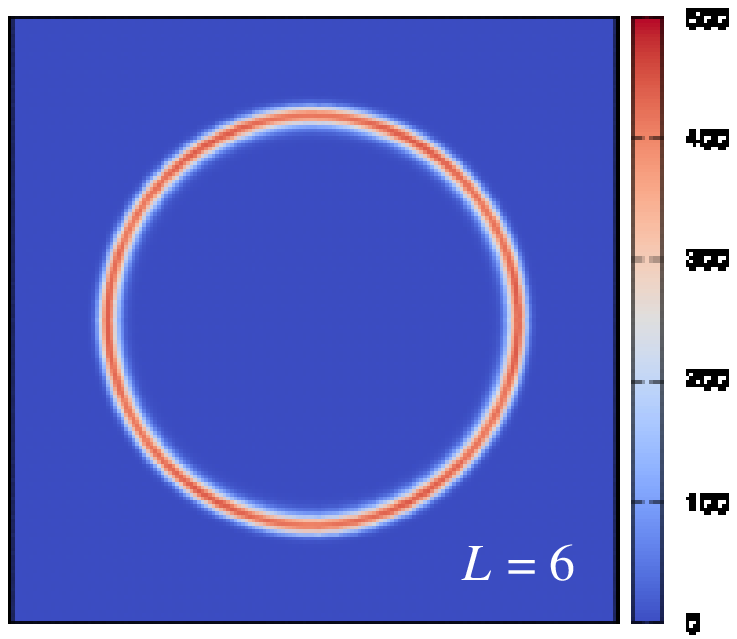}
\hspace{2.5mm}
\includegraphics[height=0.4\columnwidth]{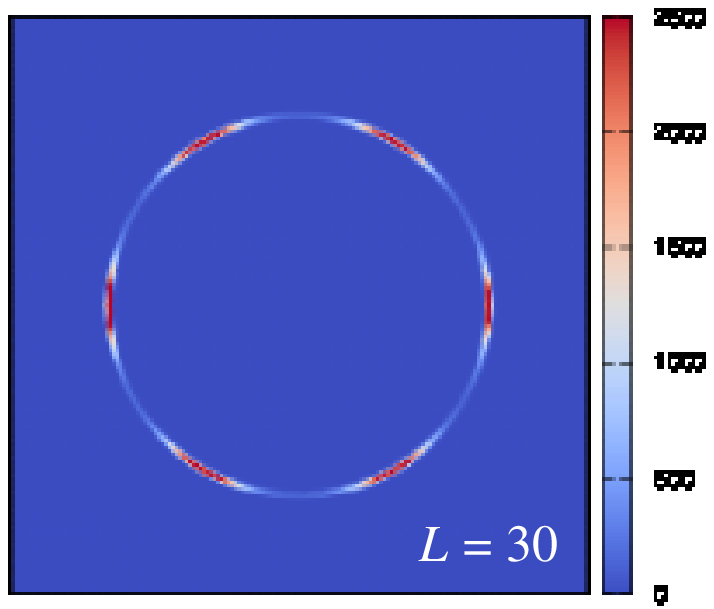}
}
\caption{(Color online)
The distribution function $P(m_x,m_y)$ for clusters with
$L=6$ (left panel) and $L=30$ (right panel) measured at $T=0.5$ ($J^a_\perp=1$).
}
\label{fig:MAP}
\end{figure}

Scaling arguments \cite{Oshikawa00,Lou07} suggest the following
finite size behavior of the two order parameters near the transition:
\begin{equation}
m = L^{-\beta/\nu}f(\tau L^{1/\nu})\,, \quad  m_6 = L^{-\beta/\nu}g(\tau L^{1/\nu_6}) \,,
\label{Mscale}
\end{equation}
where $\tau = (T_c-T)/T_c$. It is expected that $m$ and $m_6$ have the same critical exponent
$\beta$, and a leading system size dependence in the critical region, scaling as $L^{-\beta/\nu}$.
However, since the $Z_6$ anisotropy is dangerously irrelevant in 3D one expects
the scaling function to be controlled by a second divergent length scale:
While the correlation length for $m$
is $\xi \sim |\tau|^{-\nu}$ with the standard $XY$ value for $\nu$,\cite{Campostrini06}
the clock order parameter $m_6$ only becomes nonzero above a larger length scale
$\xi_6 \sim |\tau|^{-\nu_6}$  with $a_6=\nu_6/\nu>1$. This new scale can be interpreted
as a width of a domain wall between different $\psi_2$ states.
Inside the domain wall, the order parameter  angle $\theta$ smoothly varies and is not fixed
to any discrete value. Consequently, MC simulations of clusters with $L<\xi_6$ exhibit
behavior typical of a $U(1)$ symmetric model showing a finite value of $m_6$ only
once the cluster size $L$ exceeds a temperature dependent  scale $\xi_6$, see Figs.~\ref{fig:OP}(c)
and \ref{fig:OP}(d).

\begin{figure}[t]
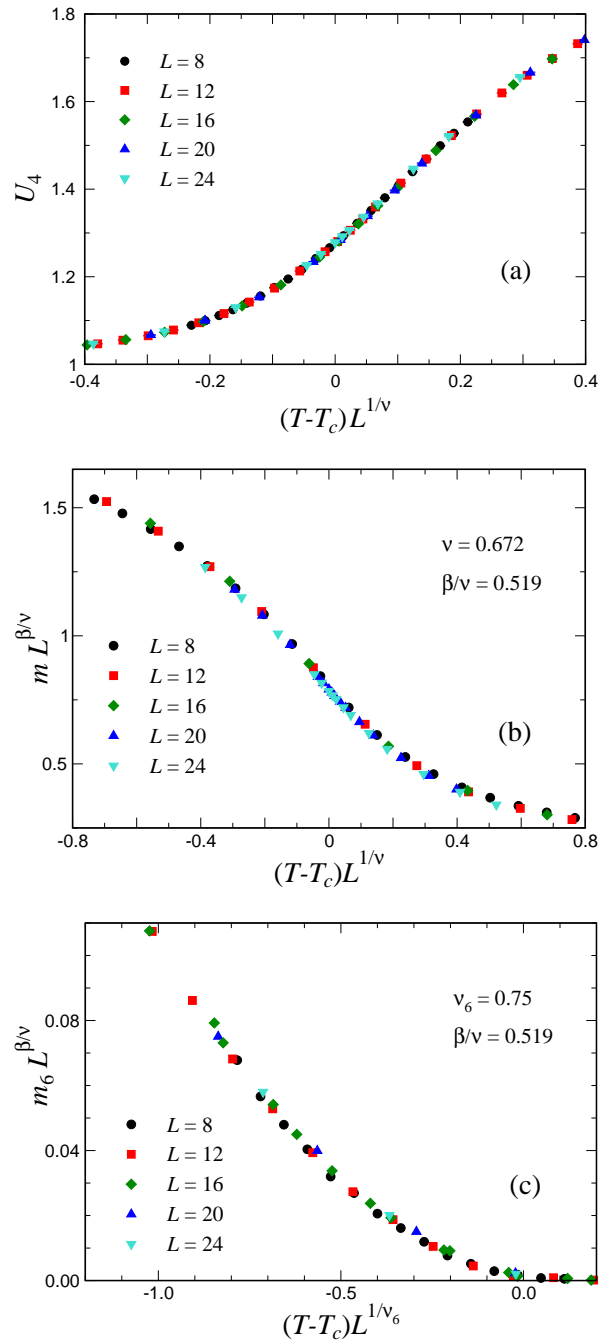

\centerline{
\includegraphics[width=0.9\columnwidth]{fig3a}
}
\vskip 4mm
\centerline{
\includegraphics[width=0.9\columnwidth]{fig3b}
}
\vskip 4mm
\centerline{
\includegraphics[width=0.9\columnwidth]{fig3c}
}
\caption{(Color online)
Finite-size scaling of the Monte Carlo data for the Binder cumulant
$U_L$ (a), the total order parameter $m$ (b), and the clock order parameter
$m_6$ (c) using  3D $XY$ critical exponents $\beta$ and $\nu$
with $T_c = 0.4454$ ($J^a_\perp=0.5$).
}
\label{fig:Scale}
\end{figure}

We illustrate the above behavior by showing in Fig.~\ref{fig:MAP} histograms for
the order parameter distribution $P(m_x,m_y)$ obtained for $J_\perp^a=1$  on
a small $L=6$ and a large $L=30$ cluster at $T=0.5$ with $T_c=0.665$. The radius
of the distribution gives the average order parameter
$\langle m\rangle$ and does not change appreciably between the two clusters.
The angular dependence of $P(m_x,m_y)$ is, however, very different in the two cases.
It is almost perfectly uniform for the small lattice, whereas
the larger system exhibits pronounced peaks corresponding
to the six domains of the $\psi_2$ state.

{\it Critical exponents.}---%
The finite-size scaling hypothesis yields $U_L= \tilde{f}(\tau L^{1/\nu})$ for
the Binder cumulant near the transition, $|\tau |\ll 1$. Hence, detailed study of
$U_L$  allows us to determine both $T_c$ and $\nu$. Using small temperature steps and
cluster sizes up to $L=24$ we have determined  $T_c = 0.4454(1)$ for $J_\perp^a = 0.5$
(see the Supplemental Material for further details\cite{SM}). As shown in
Fig.~\ref{fig:Scale}(a), an excellent data collapse is obtained for this value of $T_c$ with
$\nu=0.672$, the best estimate for the 3D $XY$ universality class. \cite{Campostrini06}

Furthermore, in Fig.~\ref{fig:Scale}(b), we find a very good collapse of the order
parameter data plotted as $mL^{\beta/\nu}$ vs.\ $\tau L^{1/\nu}$, with
$\beta/\nu=0.519$, corresponding to the 3D $XY$ model: $\beta=0.348$ and from
scaling relations, $\eta=0.038$. In the Supplemental Material we present results
of an alternative analysis, which independently confirm the estimates for $\nu$
and $\eta$ for this and other values of $J_\perp^a$.\cite{SM} Hence,
the above results unambiguously establish that the transition occurring in
the spin model for $\rm Er_2Ti_2O_7$ falls in the the 3D $XY$ universality class.

A similar scaling analysis of the MC data for $m_6$ allows us to determine
the extra exponent $\nu_6$. Using the standard value for $\beta/\nu$ we obtain
good data collapse for $\nu_6=0.75(2)$  shown in Fig.~\ref{fig:Scale}(c).
This value differs sharply from
$\nu_6\approx 1.6$ obtained by Lou {\it et al.},\cite{Lou07}
who studied the $XY$ model on a cubic lattice, perturbed by a
hexagonal anisotropy. That is, the entropically driven six-fold perturbation in the present model
appears to be more dangerous than the corresponding energy perturbation.
This is surprising, particularly so, given the observed small values for $m_6$
compared to $m$ near the transition,
which confirms the weak nature of the entropic perturbation.
Interestingly, our value of $a_6=\nu_6/\nu\approx 1.12$ appears to be closer to the value
$a_6\approx 1.3$ found for an orbital model with the same $Z_6$ symmetry.\cite{Wenzel11}
In the Supplemental Material  we present additional MC data for $J_\perp^a =1$,
which show that $\nu_6$ does not change with a varying strength of the effective
$Z_6$ anisotropy. \cite{SM} Thus, the question of universality for the critical exponent
$\nu_6$ in different realizations of 3D $Z_6$ models appears to be an open one.

{\it Neutron scattering.}---%
We now turn to the peculiar line shape for the magnetic Bragg peaks observed by Ruff {\it et al.}\
in neutron diffraction experiments on $\rm Er_2Ti_2O_7$. \cite{Ruff08}
Measuring the elastic  signal at $T=50$~mK
in the vicinity of the (2,2,0) Bragg reflection, they found that
the resolution limited peak indicative of true long-range
magnetic order coexists with broad diffuse wings characteristic of
short-range spin-liquid like correlations.
The elastic scattering can be simulated within the static approximation through the
structure factor
\begin{equation}
S({\bf q}) = \frac{1}{N} \sum_{i,j} e^{i{\bf q}\cdot({\bf r}_i-{\bf r}_j)}
\langle {\bf S}_i^\perp \cdot {\bf S}_j^\perp \rangle \,.
\end{equation}
where ${\bf S}_i^\perp$ now are spin components perpendicular to the wavevector.
Equating this to the scattering function corresponds to setting the magnetic form
factor equal to unity.
In the following we adopt the standard convention for pyrochlore antiferromagnets,
giving wavevectors in units of $2\pi/a$, where $a$ is a linear size of the cubic cell
containing 16 magnetic ions.

\begin{figure}[t]
\centerline{
\includegraphics[width=0.9\columnwidth]{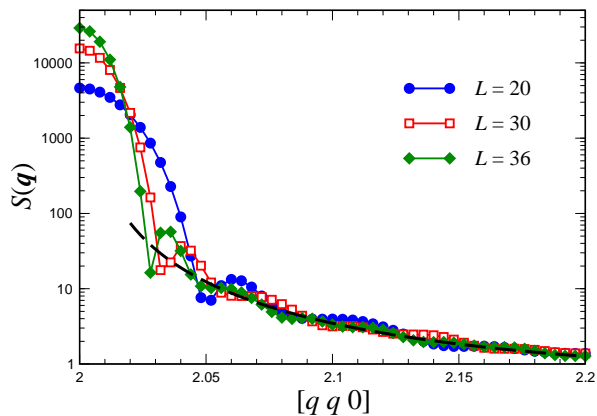}
}
\caption{(Color online)
Magnetic structure factor in the vicinity of the (2,2,0) Bragg reflection
obtained at $T=0.5$ for $J_\perp^a = 1$. The dashed line shows
$(q-2)^{-2}$ fit for the diffuse scattering wing.
}
\label{fig:Sq}
\end{figure}

Figure~\ref{fig:Sq} shows MC results for $S({\bf q})$ on a logarithmic
scale in the vicinity of the ${\bf Q}=(2,2,0)$ reflection obtained at $T=0.5$ for
$J_\perp^a=1$ on three large clusters. The Bragg peak has a width
$\propto 1/L$. After a pronounced dip,
the magnetic structure factor displays a few damped finite-size oscillations
with  period  $\Delta{\bf q}=1/L$. Besides that
$S({\bf q})$ has a finite diffuse component, which gradually decreases with the
distance from the Bragg peak.
The intensity in the scattering wing follows
a  power law  $S({\bf Q+q}) \sim 1/q^{n}$ with $n\approx 2$.
The corresponding fit is shown by a dashed line in Fig.~\ref{fig:Sq}.

This general behavior closely resembles the experimental profile of the (2,2,0)
peak  in $\rm Er_2Ti_2O_7$. \cite{Ruff08}
Additional MC simulations included in Supplemental Material
\cite{SM} show that the form of the diffuse scattering wings is maintained as the temperature
is lowered but that the intensity is reduced by a factor of 4 between $T=0.5$ and $T=0.3$
($T_c=0.636$). The persistence of the diffuse scattering is consistent with the phenomenon of
order stabilized through fluctuations, with the temperature dependence being indicative of a thermal process.
In contrast, the anomalous intensity wings have been observed experimentally
at $T=50$~mK$\ll T_N$.  We interpret this as a manifestation of the quantum
dynamics, which are present in the original effective spin-1/2 model (\ref{H0})
for  $\rm Er_2Ti_2O_7$, providing evidence that, at low temperature the order is indeed maintained
by quantum fluctuations, as has been previously proposed.
\cite{Champion03,Zhitomirsky12,Savary12}
Finally, we remark that a similar
profile for the Bragg peaks was observed in $\rm Tb_2Sn_2O_7$,\cite{Chapius07}
in which ferromagnetically aligned $z$ components of magnetic ions coexist
with antiferromagnetic ordering of transverse components.

{\it Conclusions.}---%
The model of anisotropic $XY$ pyrochlore antiferromagnet has been shown to reproduce
essentially all experimental features of the frustrated magnetic material,
$\rm Er_2Ti_2O_7$, including a second order phase transition to the six-fold symmetric
$\psi_2$ state, driven by an order by disorder mechanism.
Our extensive Monte Carlo simulations provide convincing evidence that the transition
falls into the universality class of an emergent $XY$ symmetry. The underlying
discrete symmetry manifests itself at the transition as a dangerously irrelevant
scaling variable. The scaling is found to be more dangerous than in the previously
studied case of energetic perturbation, \cite{Oshikawa00,Lou07} despite a very weak onset of
six-fold ordering in the critical region. Future work could investigate the role of
entropic forces for this apparently non-universal dangerous irrelevance by, for example,
studying a clock model with the same Hamiltonian. Simulated neutron scattering patterns
below the transition show the coexistence of both Bragg peaks and an extensive background
of diffuse scattering characteristic of spin liquid behavior in good agreement with
experiment. While in the experiment the diffuse scattering persists down to low temperature,
it weakens in our classical system providing evidence of the presence of extensive
quantum fluctuations in $\rm Er_2Ti_2O_7$.

We are grateful to M. J. P. Gingras and P. Dalmas de R\'eotier for useful discussions
and to M. V. Gvozdikova for help with Monte Carlo simulations.
MEZ and PCWH acknowledge hospitality of the Max Planck Institute for the
Physics of Complex Systems, where part of this work was performed.



\newpage
\onecolumngrid

\begin{center}
{\large\bf Supplemental material for \\ ``Nature of finite-temperature transition in anisotropic
pyrochlore Er$_\mathbf{2}$Ti$_\mathbf{2}$O$_\mathbf{7}$''}
\\
\vskip0.5cm
M. E. Zhitomirsky$^1$, P. C. W. Holdsworth$^2$, and R. Moessner$^3$\\
\vskip0.3cm
{\it \small $^1$Service de Physique Statistique, Magn\'etisme et Supraconductivit\'e,
UMR-E9001 CEA-INAC/UJF, 17 rue des Martyrs, 38054 Grenoble Cedex 9, France
\\
$^2$Laboratoire de Physique, \'Ecole Normale Sup\'erieure de Lyon,
CNRS 69364 Lyon Cedex 07, France
\\
$^3$Max-Planck-Institut f\"ur Physik komplexer Systeme, 01187 Dresden, Germany}\\
\vskip 0.2cm
{\small (Dated: February 19, 2014)}\\
\vskip 0.5cm
\end{center}

\section{Effective Spin Hamiltonian}

The rare-earth magnetic ions in pyrochlore materials are subject to a strong
crystalline electrical field with  trigonal symmetry. As a result,
the energy levels of $\rm Er^{3+}$ ions with $J=15/2$ are split
into eight Kramers doublets.
At temperatures below the crystal-field splitting energy scale
the erbium magnetic moments can be represented by effective pseudo-spin-1/2 operators.
The effective exchange Hamiltonian is a bilinear form of these operators
which is symmetric under corresponding crystal-lattice transformations.
In order to determine the most general form of anisotropic exchange interaction, let us
consider a pair of magnetic ions placed in the positions
\begin{equation}
{\bf r}_1 = (0,0,0) \quad  \textrm{and}\quad {\bf r}_2 = (1/4,1/4,0)\,.
\label{pair:suppl}
\end{equation}
The local bond frame consists of
$\hat{\bf x}_0$ pointing along the bond and $\hat{\bf z}_0$ in the direction of the
four-fold cubic axis orthogonal to the bond:
\begin{equation}
\hat{\bf x}_0 = \frac{1}{\sqrt{2}}(1,1,0)\ , \ \ \qquad
\hat{\bf y}_0 = \frac{1}{\sqrt{2}}(-1,1,0) \ , \qquad
\hat{\bf z}_0 =  (0,0,1) \ .
\end{equation}
The bond Hamiltonian is symmetric under the reflection $\hat{\sigma}_y$, $y\to -y$,
and the two-fold rotation about $\hat{\bf z}_0$, $C^2_z$: $x\to -x$, $1\leftrightarrow 2$.
As a result,
\begin{equation}
\hat{\cal H}_{12} =  J_{x_0x_0} S_1^{x_0}S_2^{x_0} + J_{y_0y_0} S_1^{y_0}S_2^{y_0} +
J_{z_0z_0} S_1^{z_0}S_2^{z_0} + D (S_1^{z_0}S_2^{x_0} - S_1^{x_0}S_2^{z_0} ) \ .
\label{H120}
\end{equation}
Thus, the bond Hamiltonian is parameterized by four independent interaction
constants, cf.~[1]. The first three terms in (\ref{H120})
are symmetric under permutation of spins and describe the anisotropic
exchange interactions, whereas the last antisymmetric term corresponds to
the Dzyaloshinskii-Moriya (DM) interaction.
In agreement with Ref.~[2], the DM vector for a given bond
is parallel to the opposite bond of the same tetrahedron.

For pyrochlore materials with pronounced Ising or planar anisotropy
it is convenient to transform to   a local basis with $\hat{\bf z}_i$
being oriented along the trigonal axis on each site.
For the pair of spins (\ref{pair:suppl}), a suitable choice of the local axes is
\begin{eqnarray}
&& \hat{\bf x}_1 = \frac{1}{\sqrt{6}}(1,1,-2)\ , \ \ \qquad
\hat{\bf y}_1 = \frac{1}{\sqrt{2}}(-1,1,0) \ , \qquad
\hat{\bf z}_1 = \frac{1}{\sqrt{3}}(1,1,1) \ , \nonumber \\
&& \hat{\bf x}_2 = \frac{1}{\sqrt{6}}(-1,-1,-2)\ , \quad
\hat{\bf y}_2 = \frac{1}{\sqrt{2}}(1,-1,0)\ , \qquad
\hat{\bf z}_2 = \frac{1}{\sqrt{3}}(-1,-1,1) \ .
\label{x12}
\end{eqnarray}
The bond Hamiltonian (\ref{H120}) transforms into
\begin{equation}
\hat{\cal H}_{12} =  J_{xx} S_1^{x}S_2^{x} + J_{yy} S_1^{y}S_2^{y} +
J_{zz} S_1^{z}S_2^{z} + J_{xz} (S_1^{z}S_2^{x} + S_1^{x}S_2^{z} )
\label{H121}
\end{equation}
in the new basis. Due to staggered local axes, the antisymmetric DM term acquires a symmetric form in this basis.
Extending (\ref{H121}) over the whole lattice we write, following our previous work [3]
\begin{eqnarray}
&& \hat{\cal H} =  \sum_{\langle ij\rangle}\Bigl\{ J_{zz} S_i^zS_j^z +
J_\perp {\bf S}^\perp_i\cdot{\bf S}^\perp_j + J_\perp^a
({\bf S}^\perp_i\cdot \hat{\bf r}_{ij})({\bf S}^\perp_j\cdot \hat{\bf r}_{ij})
 + J_{z\perp}
\bigl[S_j^z({\bf S}^\perp_i\cdot \hat{\bf r}_{ij}) + S_i^z({\bf S}^\perp_j\cdot \hat{\bf r}_{ji})\bigr]
\Bigr\}\,,
\label{Hmzh}
\end{eqnarray}
where ${\bf S}^\perp_i$ are spin components perpendicular to the
local trigonal axes $z_i$ and $\hat{\bf r}_{ij}$ is a unit vector in the bond direction.
Using the coupling constants defined by Eq.~(\ref{Hmzh}), the bond Hamiltonian for a pair of spins (\ref{pair:suppl}) in the basis (\ref{x12})
can be once more expressed as
\begin{eqnarray}
&& \hat{\cal H}_{12} = J_{zz} S_1^z S_2^z - J_\perp S_1^y S_2^y   + \frac{1}{3} (J_\perp - J_\perp^a) S_1^x S_2^x
+ \frac{1}{\sqrt{3}} J_{z\perp} (S_1^z S_2^x + S_1^x S_2^z)\ .
\label{H12}
\end{eqnarray}

An alternative form for the effective Hamiltonian for a spin-1/2 anisotropic pyrochlore
antiferromagnet has been used in [4-6]:
\begin{eqnarray}
\hat{\cal H} & = & \sum_{\langle ij\rangle}\bigl\{ J_{zz} S_i^zS_j^z -
J_\pm (S^+_i S^-_j + S_i^-S_j^+) + J_{\pm\pm} (\gamma_{ij}S^+_i S^+_j + \gamma^*_{ij}S_i^-S_j^-)
+ J_{z\pm} \bigl[S_j^z(\zeta_{ij}S^+_i +\zeta^*_{ij}S_i^-)  + i \leftrightarrow j\bigr]
\bigr\}\,,
\label{Hsav}
\end{eqnarray}
where $\gamma_{ij} = e^{2\pi in/3}$ and $\zeta_{ij} = -\gamma^*_{ij}$
depending on the bond direction. For the axes choice (\ref{x12}) one has
$\gamma_{12} = -\zeta_{12}=1$. Substituting this into Eq.~(\ref{Hsav}) and
comparing  to (\ref{H12})  we obtain the following relation between the two sets
of exchange parameters:
\begin{equation}
J_\perp = 2(J_\pm + J_{\pm\pm})\,, \qquad
J^a_\perp = 8J_\pm - 4J_{\pm\pm}\,, \qquad
J_{z\perp} = -2\sqrt{3}J_{z\pm}\,.
\label{Jconv}
\end{equation}

Using  inelastic neutron-scattering  in strong magnetic field,
Savary {\it et al.}~[6]
estimated the exchange constants in $\rm Er_2Ti_2O_7$ as
\begin{equation}
J_\pm = 6.5\pm0.75\ , \quad
J_{\pm\pm} = 4.2\pm0.5\ , \quad
J_{zz} = -2.5\pm1.8\ , \quad
J_{z\pm} = -0.88\pm1.5~\ , \qquad \textrm{in}\ 10^{-2}\ \textrm{meV}\ .
\end{equation}
Applying (\ref{Jconv}) we obtain
\begin{equation}
J_\perp = 0.21(2)~\textrm{meV}\ , \quad J^a_\perp = 0.35(5)~\textrm{meV}\ , \quad
J_{zz} = -0.025(2)~\textrm{meV}\ , \quad J_{z\perp} = 0.03(5)~\textrm{meV}\ .
\end{equation}
The above values confirm the planar character of the interaction
between Er$^{3+}$ moments as well as a significant anisotropy for the in-plane exchange
constants $J_\perp^a/J_\perp \approx 1.67$. Our previous estimate [3]
of the main exchange  parameters for $\rm Er_2Ti_2O_7$ based on zero-field magnon
dispersion data by Ruff \textit{et al}.~[7] gives
$J_\perp \approx 0.24$~\textrm{meV} and $J^a_\perp/J_\perp \approx 1.2$.
There is a reasonable correspondence between the two sets of parameters.
Hence, the gross features in the behavior of
$\rm Er_2Ti_2O_7$ can be explained in the model with just two exchange
parameters $J_\perp$ and $J^a_\perp$.

\section{Monte Carlo results}

\begin{figure}[b]
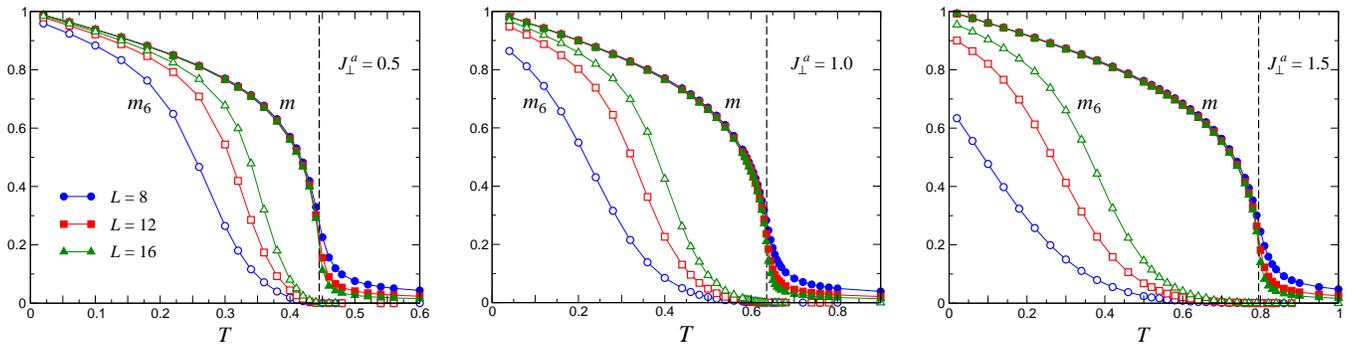

\centerline{
\includegraphics[height=0.25\columnwidth]{OP_0.5.eps} \hspace{3mm}
\includegraphics[height=0.25\columnwidth]{OP_1.0.eps} \hspace{3mm}
\includegraphics[height=0.25\columnwidth]{OP_1.5.eps}
}
\caption{Temperature dependence of the two order parameters
on clusters with  $L = 8,12,16$ for three different values of $J_\perp^a$.
Full symbols correspond to $m$, while open symbols show $m_6$.
The vertical dashed lines indicate the respective transition temperatures
determined from the crossing points of the Binder cumulants $U_L$.
}
\label{fig:OPs}
\end{figure}

Classical Monte Carlo (MC) simulations have been  performed for the
$XXZ$ spin Hamiltonian
\begin{equation}
\hat{\cal H} = \sum_{\langle ij\rangle} \Bigl[ {\bf S}^\perp_i\cdot{\bf S}^\perp_j +
J_\perp^a ({\bf S}^\perp_i\cdot \hat{\bf r}_{ij})({\bf S}^\perp_j\cdot \hat{\bf r}_{ij}) \Bigr]
\label{H}
\end{equation}
using the hybrid algorithm described in the main text.

Temperature dependence of the two relevant order parameters for three
values of $J^a_\perp$ is shown in Figure \ref{fig:OPs}. While the total order parameter
$m$ exhibits a behavior typical of a second-order transition, the clock order parameter
$m_6$ has an anomalous temperature dependence with a remarkable
inverse finite-size scaling. Such  behavior is explained by the dangerously  irrelevant role
of the six-fold anisotropy at a 3D $XY$ transition and appearance
of an additional length-scale $\xi_6>\xi$ below $T_c$.
The temperature interval, where $m_6$ is suppressed compared to $m$, grows with
increasing $J_\perp^a$. This tendency is explained by a reduction of the effective
effective $Z_6$ anisotropy generated by thermal as well as by quantum fluctuations
[3]. Indeed, for $J_\perp^a=4$ the bond Hamiltonian (\ref{H12})
acquires a spurious symmetry between $x$ and $y$ components such that the degeneracy
between $\psi_2$ and $\psi_3$ states becomes exact [8].

Below we provide additional Monte Carlo data and their analysis, which further support
the conclusions about the critical behavior of the model (\ref{H})
presented in the main text.

\begin{figure}[t]
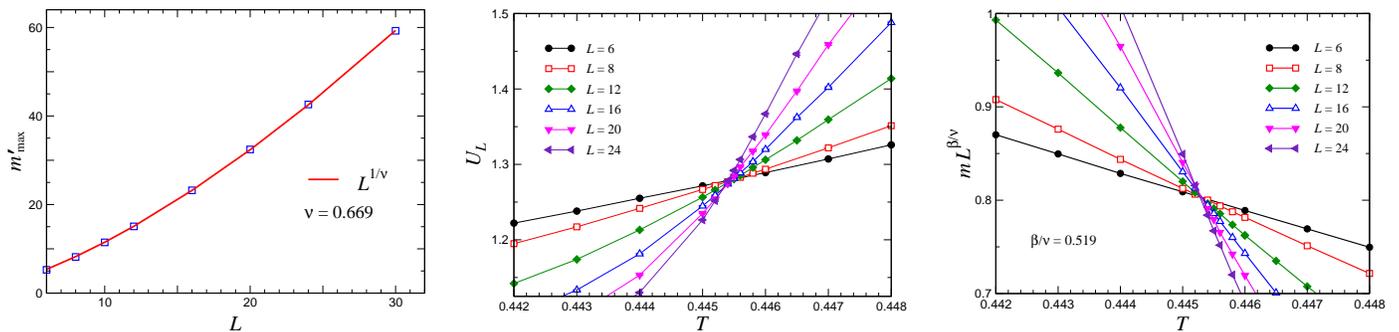

\centerline{
\includegraphics[height=0.24\columnwidth]{dermax_0.5.eps} \hspace{3mm}
\includegraphics[height=0.24\columnwidth]{binder_0.5.eps} \hspace{3mm}
\includegraphics[height=0.24\columnwidth]{OP_0.5crossing.eps}
}
\caption{Monte Carlo results for the model (\ref{H}) with $J_\perp^a=0.5$.
Left panel: finite-size scaling of the maximum of the logarithmic derivative of the order
parameter $m'_{\rm log}$, see Eq.~(\ref{mlog}). The solid line is a fit to  $L^{1/\nu}$,
with $\nu \approx 0.669$.
Central panel: temperature dependence of the Binder cumulants close to $T_c$.
Right panel: temperature dependence of the rescaled order parameter.
}
\label{fig:OPmax}
\end{figure}

To verify the 3D $XY$ universality of the transition
we fix the exchange anisotropy parameter to $J_\perp^a=0.5$, the lowest value  among three
plots  in Fig.~\ref{fig:OPs}. The entropically generated $Z_6$  anisotropy is strongest and,
hence, most relevant for this value of  $J_\perp^a$. To determine the value of the critical
exponent $\nu$ we measure, in Monte Carlo simulations, the logarithmic derivative of the order
parameter
\begin{equation}
m'_{\rm log} = \frac{1}{m}\,\frac{d m}{dT} = = \frac{1}{T^2}\biggl(
\frac{\langle m\hat{\cal H}\rangle}{\langle m\rangle} - \langle \hat{\cal H}\rangle\biggr)\,.
\label{mlog}
\end{equation}
As a function of temperature, $m'_{\rm log}$ exhibits a maximum near $T_c$.
The height of the maximum
$m'_{\rm max} = \max\{m'_{\rm log}\}$ scales with  system size according to
\begin{equation}
m'_{\rm max}  \sim L^{1/\nu}\,.
\end{equation}
Thus, from the scaling of the peak data one can obtain an unbiased estimate for the critical
exponent $\nu$. The corresponding Monte Carlo results are presented in the left panel of
Figure~\ref{fig:OPmax}.
The best fit is shown by the solid line and yields $\nu = 0.669(2)$, which nicely agrees
with the 3D $XY$ value $\nu = 0.672$ [9].

To fully establish the universality class of the transition, we need to evaluate
a second critical exponent, for example,  $\eta$. All other critical exponents
can be determined using known values of $\eta$ and $\nu$ from the scaling relations.
In particular, the order parameter exponent $\beta$ is found from
\begin{equation}
\frac{\beta}{\nu} =  \frac{d}{2} - 1 + \frac{\eta}{2}
\end{equation}
with $\beta/\nu = 0.5+\eta/2$ in 3D. Thus, we shall be checking the value of $\beta/\nu=0.519$
rather than the value of $\eta=0.038$ for the 3D $XY$ universality class [9].

The ratio $\beta/\nu$ enters the finite-size scaling law for $m$:
\begin{equation}
m = L^{-\beta/\nu} f(\tau L^{1/\nu})
\end{equation}
with $\tau = T-T_c$, see, {\it e.g.}, [10].
Hence, by plotting $m L^{\beta/\nu}$ versus $T$ for various system sizes
with the correct value of $\beta/\nu$ one should obtain crossing
of different curves at $T=T_c$, very similar to the standard procedure widely adopted
for Binder cumulants.

Before using the above approach for the rescaled order parameter one needs to obtain
an independent estimate for the transition temperature.
The central panel of Fig.~\ref{fig:OPmax} shows the Binder cumulants
$U_L(T)=\langle m^4\rangle/\langle m^2\rangle^2$ close to $T_c$.
From the crossing point we obtain an accurate value of the critical temperature $T_c^U \approx 0.4454(1)$.
The right panel of the same Figure presents the MC data for the scaled order parameter
$m L^{\beta/\nu}$ with $\beta/\nu= 0.519$ corresponding to the 3D $XY$ value of $\eta=0.038$, see
[9]. The curves corresponding to different linear sizes $L$
cross at $T_c^m \approx 0.4453(1)$. The remarkable closeness of the two estimates for the transition
temperature, $T_c^U$ and $T_c^m$, may be used as a quantitative confirmation
of the 3D $XY$ value for $\eta$.

\begin{figure}[t]
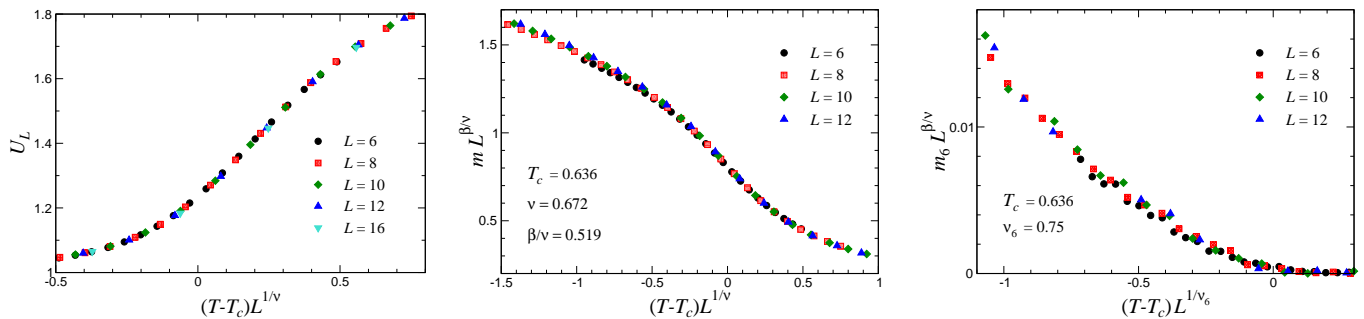

\centerline{
\includegraphics[height=0.23\columnwidth]{binder_1.0scaling.eps}\hspace{3mm}
\includegraphics[height=0.23\columnwidth]{OP_1.0scalingM.eps} \hspace{3mm}
\includegraphics[height=0.23\columnwidth]{OP_1.0scalingM6.eps}
}
\caption{
Monte Carlo results for the model (\ref{H}) with $J_\perp^a=1$.
Finite-size scaling of the Binder cumulants
$U_L$ (left panel), the total order parameter $m$ (central panel), and
the clock order parameter $m_6$ (right panel) using the 3D $XY$ critical exponents
$\beta$ and $\nu$ with $T_c = 0.636$.
}
\label{fig:OPscaling}
\end{figure}

To conclude the analysis, we give a brief check of the critical behavior for another value
of $J_\perp^a = 1$. Figure~\ref{fig:OPscaling} shows  standard data collapse
plots for the Binder cumulants (left panel) and the order parameter (central panel)
obtained with the 3D  $XY$ values of $\nu$ and $\beta$ and $T_c = 0.636$.
This perfect data collapse further confirms the expected irrelevance of the anisotropy
in 3D. A similar scaling analysis of the MC data for the clock order parameter $m_6$
is shown in the right panel.
A good data collapse is obtained with $\nu_6=0.75(2)$, the same value as for
$J_\perp^a = 0.5$ (see the main text). A somewhat larger statistical scattering of
the MC data for $J_\perp^a = 1$ compared to $J_\perp^a = 0.5$ (main text, Fig.~3(c))
appears because of the much smaller values of $m_6$
in the vicinity of $T_c$, see Fig.~\ref{fig:OPs}.
Thus, the critical properties of the clock order parameter appear
to be independent of the strength of the anisotropic exchange $J_\perp^a$.

\begin{figure}[b]
\centerline{
\includegraphics[height=0.3\columnwidth]{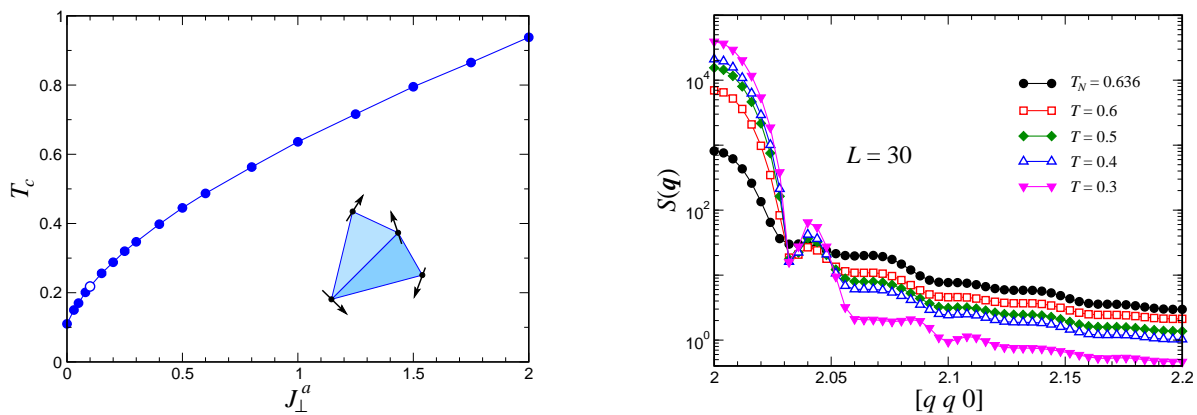}\hspace{15mm}
\includegraphics[height=0.297\columnwidth]{sq_compT.eps}
}
\caption{
Left panel: Dependence of the transition temperature $T_c$ versus the anisotropy
parameter $J_\perp^a$ (both in units of $J_\perp$)
for the anisotropic $XY$ pyrochlore (\ref{H}).
Right panel: Temperature variations of the magnetic structure factor in the vicinity of
the (2,2,0) Bragg reflection obtained on  a cluster with the linear size $L=30$
($J_\perp^a = 1$).
}
\label{fig:Sqs}
\end{figure}

To finish the discussion of the transition in the anisotropic $XY$ pyrochlore
antiferromagnet (\ref{H}) we show in the left panel of Fig.~\ref{fig:Sqs} the dependence
of $T_c$ on the anisotropy parameter $J_\perp^a$. The open circle shows an approximate
position of the tricritical point $(J^*,T^*) \approx (0.1,0.218)$.
For $0<J_\perp^{a}<J^*$ the transition into the $\psi_2$ state is of the first order,
whereas for $J_\perp^{a}>J^*$ it becomes to be a second order transition.

Finally, we show in the right panel of Figure \ref{fig:Sqs} the temperature dependence of the diffuse
scattering intensity in the vicinity of the (2,2,0) magnetic Bragg peak.
Our classical Monte Carlo results  for $S({\bf q})$ strongly  resemble the experimental neutron data
of Ruff {\it et al.}~[7] on the coexistence of
static magnetic order and the diffuse magnetic component in $\rm Er_2Ti_2O_7$.
For the classical model (\ref{H}) the shape of the diffuse wing stays approximately
constant with $S({\bf q}) \approx 1/(q-2)^2$ power law decay, whereas the intensity is significantly
reduced with lowering $T$.  We interpret the persistent presence
of the diffuse scattering below $T_N$ as being due to  short-wavelength fluctuations, which are
responsible for the order by disorder selection in the anisotropic $XY$ pyrochlore antiferromagnet
(\ref{Hmzh}).

\newpage

\vskip0.5cm

\begin{center}
\rule{0.4\linewidth}{0.35mm}
\end{center}

\vskip0.5cm

\noindent
[1]\ \
S. H. Curnoe, Phys. Rev. B \textbf{78}, 094418 (2008).
\vskip1mm

\noindent
[2]\ \
M. Elhajal, B. Canals, R. Sunyer, and C. Lacroix,
Phys. Rev. B \textbf{71}, 094420 (2005).
\vskip1mm

\noindent
[3]\ \
M. E. Zhitomirsky, M. V. Gvozdikova, P. C. W. Holdsworth,
and R. Moessner, Phys. Rev. Lett. {\bf 109}, 077204 \\ \hspace*{5mm} (2012).
\vskip1mm

\noindent
[4]\ \
S. Onoda and Y. Tanaka, Phys. Rev. Lett. \textbf{105}, 047201 (2010).
\vskip1mm

\noindent
[5]\ \
K. A. Ross, L. Savary, B. D. Gaulin, and L. Balents,
Phys. Rev. X \textbf{1}, 021002 (2011).
\vskip1mm

\noindent
[6]\ \
L. Savary, K. A. Ross, B. D. Gaulin, J. P. C. Ruff,
and L. Balents, Phys. Rev. Lett. {\bf 109}, 167201  (2012).
\vskip1mm

\noindent
[7]\ \
J. P. C. Ruff, J. P. Clancy, A. Bourque, M. A. White, M. Ramazanoglu,
J. S. Gardner, Y. Qiu, J. R. D. Copley, \\ \hspace*{5mm}
M. B. Johnson, H. A. Dabkowska,
and B. D. Gaulin,
Phys. Rev. Lett. {\bf 101}, 147205 (2008).
\vskip1mm

\noindent
[8]\ \
A. W. C. Wong, Z. Hao, and M. J. P. Gingras,
Phys. Rev. B \textbf{88}, 144402 (2013).
\vskip1mm

\noindent
[9]\ \
M. Campostrini, M. Hasenbusch, A. Pelissetto, and E. Vicari,
Phys. Rev. B \textbf{74}, 144506 (2006).
\vskip1mm

\noindent
[10]\
M. N. Barber, Finite-size scaling, in \textit{Phase Transitionas and Critical
Phenomena} vol.~8, edited by C. Domb and \\ \hspace*{5mm} J. L. Lebowitz (Academic Press, London, 1983).

\end{document}